\documentclass[nofootinbib,showpacs,preprintnumbers,amsmath,amssymb,floatfix]
{revtex4-1}
\usepackage{epsf}

\begin{document}


\title{COMMENT ON THE FROZEN QCD COUPLING }

\vspace*{0.3 cm}

\author{B.I.~Ermolaev}
\affiliation{Ioffe Physico-Technical Institute, 194021
 St.Petersburg, Russia}
\author{M.~Greco}
\affiliation{Department of Physics and INFN, University Roma Tre,
Rome, Italy}
\author{S.I.~Troyan}
\affiliation{St.Petersburg Institute of Nuclear Physics, 188300
Gatchina, Russia}

\begin{abstract}
The frozen QCD coupling is a parameter often used as an effective fixed coupling.
It is supposed to mimic  both the running coupling effects
and the lack of knowledge of $\alpha_s$ in the infrared region. Usually the value of the frozen coupling
is fixed from the analysis of the experimental data.
We present a novel way to define such coupling(s) independently of the experiments. We
argue that there are different frozen couplings which are used in
the double- and single- logarithmic approximations.
We introduce three kinds of the frozen couplings:
the coupling used in DLA with a time-like argument (i.e. the coupling present in the non-singlet scattering
amplitudes and DIS structure functions) which we find
 $0.24$ approximately; the DLA coupling with a space-like argument ( $e^+e^-$ -annihilation, in DY
processes and in any scattering amplitude in the hard or backward kinematics) which is a factor two larger, namely $0.48$. We also show that
the frozen coupling in the
single-logarithmic evolution equations like BFKL can be defined with much less accuracy compared to DLA,
and our estimate for this coupling is $0.1$. Our estimates for the singlet and non-singlet
intercepts are also in a good agreement with the results available in the literature.
\end{abstract}


\maketitle

\section{Introduction}

The concept of a frozen QCD coupling was introduced long ago (see e.g. Ref.~\cite{greco})
and is still quite popular nowadays (see e.g. Refs.~\cite{webb}- \cite{zotov}). The origin of
the term comes up with the divergent infrared behavior of the well-known RG -expression for
$\alpha_s$: the frozen coupling is a constant and therefore it can be used in the IR domain.
Another motivation for introducing the frozen coupling is
that the perturbative QCD coupling is running, and accounting for the running $\alpha_s$ effects should have been done in
every calculation. However, this makes some QCD calculations quite complicated,
so for rough estimates, it might be  convenient to use some
effective coupling which mimics the running of $\alpha_s$ in the perturbative region and
also can be used beyond it.
The values of the frozen couplings are usually fixed from purely phenomenological
considerations  to get an agreement with experimental data. Quite often the
frozen coupling is used in combination with other phenomenological parameters to
describe hadronic reactions (e.g. parton distributions, etc), which
makes more involved fixing its value.
A fixed $\alpha_s$ has often been used in various calculations done in the
framework of the Leading Logarithmic Approximation (with the leading contributions
being either double- or single- logarithms, depending on the situation) where the most
important logarithmic contributions are totally resummed, while $\alpha_s$
is treated as a fixed parameter and its argument is set a posteriori
from physical considerations.

In contrast, in the present paper we suggest a novel way
to define the fixed (frozen) coupling, which does not involve any analysis of experimental data and
does not coincide with the often used  generalization of the DGLAP-parametrization
$\alpha_s (Q^2)$ to $\alpha_s = \alpha_s (\mu^2 + Q^2)$. We begin by
considering the parametrization of $\alpha_s$ in the evolution equations. In particular,
they could be the DGLAP equations\cite{dglap} or the generalizations of them\cite{acta} to
the small-$x$ region, or the BFKL equation\cite{bfkl,bfklnlo}. We show how the effective running coupling present in the
equations can be replaced by a constant
coupling fixed at some scale both in the double-logarithmic (DLA) and single-logarithmic (SLA) Approximation.
In order to set the scales in DLA, we use the Principle of Minimal Sensitivity
(PMS) suggested in Ref.~\cite{pms}
and then applied to various QCD problems in Ref.~\cite{pms1}. PMS has also proved to be efficient
in the analysis of electro-production of vector mesons\cite{iv} and for setting the scale of the BFKL
Pomeron\cite{kim}. Nevertheless, as PMS does not have profound theoretical grounds, we check our estimates
for the frozen DL couplings obtained with PMS. By doing so, we avoid any comparison of our results
to experimental data because the latter usually
involve a set of phenomenological parameters. Instead, we compare our estimates for the intercepts
of the DIS structure functions
to the available results obtained earlier without using PMS. Considering the scales for the SL frozen
couplings, we set the SL scales first without using PMS and than applying it. Then, we compare these results
to each other and
to the result obtained in Ref.\cite{kim} in the context of analysis of the BFKL Pomeron
intercept.
We also stress that setting the frozen couplings in SLA
is more involved and  much less reliable compared to DLA.

Our paper is organized as follows: in Sect.~II we briefly remind what is the parametrization
of $\alpha_s$ in the Born amplitudes and in the evolution equations of the Bethe-Salpeter
type. In Sect.~III we provide the general grounds for keeping the QCD coupling fixed. In Sect.~IV we
apply these results to fix the DL frozen couplings. Then, in Sect.~V we estimate
the SL couplings. In Sect.~VI we check our estimates for the frozen DL couplings, using the
couplings to calculate the intercepts of the non-singlet DIS structure functions and compare our
estimates to the results available in the literature. In Sect.~VII we compare our
estimates for the frozen SL couplings to the available results. We make a brief
comment on the use of the standard parametrization of the QCD coupling in Sect.~VIII.
Finally, Sect.~VIII is for concluding
remarks.

\section{Running QCD coupling in the scattering amplitudes}

In this Sect. we remind the results\cite{etalpha} on the parametrization of the running $\alpha_s$.
The rigorous knowledge on the QCD coupling comes from the RG -equations. In the LO approximation,
the coupling at $q^2 < 0$ is

\begin{equation}\label{alphaqcd}
\alpha_s (q^2) = \frac{1}{b \ln \left(|q^2|/\Lambda^2\right)} = \frac{1}{b \ln \left(-q^2/\Lambda^2\right)},
\end{equation}
with $b = (11N - 2n_f)/(12 \pi)$. Eq.~(\ref{alphaqcd}) is valid for $|q^2| \gg \Lambda^2$.
However, in different scattering amplitudes the effective coupling  can
be given by more involved expressions than Eq.~(\ref{alphaqcd}).
In the first place  we consider the scattering amplitude for
the forward annihilation of a quark-antiquark pair  into another pair:
$q (p_1) ~\widetilde{q}(p_2) \to q'(p'_1) ~\widetilde{q}'(p'_2)$ in the forward kinematics  where

\begin{equation}\label{fkinem}
s = (p_1 + p_2) \gg -t = -(p'_1 - p_1)^2  \approx 0.
\end{equation}

\subsection{Parametrization of the QCD coupling in the Born amplitudes}

In the Born approximation the scattering amplitude $B$ for this process is represented by
the single Feynman graph depicted in Fig.~1, where the intermediate gluon propagates in the $s$-channel,
and is given by the following expression (with the color structure dropped as unessential):

\begin{figure}[h]
\begin{center}
\begin{picture}(160,120)
\put(0,0){ \epsfbox{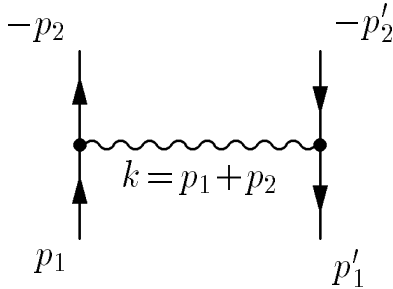} }
\end{picture}
\end{center}
\caption{\label{frozenfig1} Born amplitude for $q\bar{q}$ -annihilation.}
\end{figure}

\begin{equation}\label{born}
B = \frac{\bar{u}_2\gamma_{\lambda}u_1 \bar{u}'_1\gamma_{\mu}u'_2}{s}~d_{\lambda \mu}~ B_q,
\end{equation}
where $B_q$ is the invariant Born amplitude:
\begin{equation}\label{bq1}
B_q = -
\alpha_s \frac{s}{s + \imath \epsilon}.
\end{equation}
The coupling $\alpha_s$ in Eq.~(\ref{bq1}) is a constant. Accounting for the
radiative corrections converts the Born amplitude $B$ into a high-order amplitude where
a part of the corrections is attributed to the QCD coupling.
When the total resummation of the radiative corrections has been done,
the Born couplings are converted into the running ones.
In order to apply the Born amplitudes to physical processes, the following scenario
is often used. Accounting for the radiative corrections is
done in two steps: \\
\textbf{(i)} Total resummation of the corrections to the Born $\alpha_s$. This converts the fixed
coupling of Eq.~(\ref{bq1}) into the running one, leaving the rest of (\ref{bq1}) unchanged.  \\
\textbf{(ii)} Accounting for the rest of  radiative corrections.\\
Let us notice that for self-consistency both step \textbf{(i)} and step \textbf{(ii)} should
involve the total resummation of the radiative corrections at least with logarithmic
accuracy. However, in practice step  \textbf{(ii)} is often done to a fixed order in the coupling.
After the step \textbf{(i)}  has been done, the amplitude
$\tilde{B}_q$ in  Eq.~(\ref{bq1}) is replaced by

\begin{equation}\label{bq2}
B_q =  \alpha_s (s) \frac{s}{s  + \imath \epsilon} .
\end{equation}

The amplitude $B_q$ is commonly referred as the Born amplitude and we will follow
this terminology throughout the present paper. Now we focus on the coupling in Eq.~(\ref{bq2}).
Being defined by the RG equations for space-like momenta $q$, the QCD coupling
acquires an imaginary
part when its argument $s$ is time-like.
%

\subsection{QCD coupling in evolution equations}

The expressions for the scattering amplitudes and DIS structure functions are usually obtained by solving
evolution equations. Such equations are mostly of the Bethe-Salpeter type.
They are often constructed in such a way
that their r.h.s. include the Born
contribution and a convolution where one gluon is factorized:
\begin{equation}\label{bseq}
M = M_{Born} + M \otimes M_{Born} .
\end{equation}

In particular, the term $A = M \otimes M_{Born}$ is present in the DGLAP equations\cite{dglap} in the integral form
and similar terms are present in the generalizations of DGLAP  to the small-$x$ region (see the overviews\cite{acta}).
The evolution equations of the Bethe-Salpeter type are used to describe various processes. In order to be more specific,
we consider in the present paper the convolution term $A$ contributing to the DGLAP evolution
equation for the DIS structure functions. In this regard we follow Ref.~\cite{etalpha}.
The term $A$ is shown in Fig.~2.

\begin{figure}[h]
\begin{center}
\begin{picture}(125,160)
\put(0,0){ \epsfbox{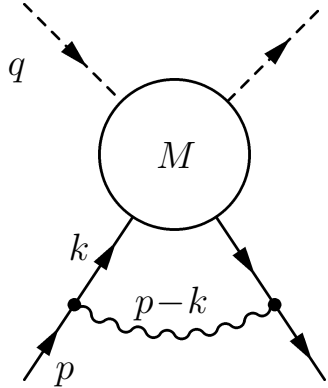} }
\end{picture}
\end{center}
\caption{\label{frozenfig2} The integral contribution to
Eq.~(\ref{bseq}). The $w$ -cut is implied, though not shown
explicitly.}
\end{figure}

The graph depicted in Fig.~2 implies that the leading small-$x$ contributions are
double-logarithmic. The graph includes the blob $M$ corresponding to the scattering between the virtual photon
with momentum $q$ and the quark with momentum $k$. Besides, it includes
the gluon with momentum $m= p-k$ factorized out of the blob $M$. This gluon propagates in the $s$-channel.
After simplification of the spinor structure of the numerator, the second term in the r.h.s. of Eq.~(\ref{bseq}) is

\begin{equation}\label{bsa}
A (x, Q^2) = \int d \beta d k_{\perp}^2 \int_{0}^{w}
d m^2 M \left(x, Q^2, w\beta, (\beta m^2 + k_{\perp}^2)\right)
\frac{1}{(m^2 \beta + k_{\perp}^2)} \Im \left[\frac{\alpha_s (m^2)}{(m^2 + \imath \epsilon)}\right],
\end{equation}

where we have used the standard notations $w=2pq, Q^2 = - q^2, x = Q^2/w$. The momenta of the intermediate quark is
parameterized by the Sudakov variables $\beta, m^2, k_{\perp}$:

\begin{equation}\label{sud}
k_{\lambda} = \beta (q + x p)_{\lambda} - (m^2/w) p_{\lambda} + k_{\perp~\lambda}
\end{equation}
so that
\begin{equation}\label{k2}
(p - k)^2 = m^2, k^2 = - \frac{m^2 \beta + k_{\perp}^2}{1- \beta},
(q + k)^2 = w\beta -  x\frac{m^2 \beta + k_{\perp}^2}{1- \beta}  -Q^2 + k^2 .
\end{equation}

The integration over $m^2$ in Eq.~(\ref{bsa}) involves the unknown blob $M$, so it cannot be done
straightforwardly. However, it can be done approximately, when the most essential contributions in
$M$ comes from the region where
\begin{equation}\label{kbm}
m^2 \beta \ll k_{\perp}^2,
\end{equation}
as usually occurs when the ladder graphs are calculated with the logarithmic accuracy. In this case the upper
limit of integration over $m^2$ is replaced by $k_{\perp}^2/\beta$, after that $M$ is
independent of $m^2$, so the integration over $m^2$ can be performed. This makes possible
 to represent  $A(x,Q^2)$ as a convolution of $M$ and an effective coupling $\alpha_{eff}$:
\begin{equation}\label{bsaeff}
A (x, Q^2) = \int d \beta \frac{d k_{\perp}^2} {k_{\perp}^2}
M \left(x, Q^2, w\beta, k_{\perp}^2)\right) \alpha_{eff} ,
\end{equation}
with
\begin{eqnarray}\label{aeff}
\alpha_{eff} &=& \frac{1}{b}\int_0^{k_{\perp}^2/\beta}
d m^2 \delta (m^2 - \mu^2)
 \frac{t}{(t^2 + \pi^2)} - \frac{1}{b}\int_{\mu^2}^{k_{\perp}^2/\beta}
\frac{d m^2}{m^2} \frac{1}{(t^2 + \pi^2)}  \\
\nonumber
&=& \frac{1}{b} \frac{l_0}{(l^2_0 + \pi^2)} - \frac{1}{\pi b} \arctan \left(\frac{\pi}{l_0}\right)
+  \frac{1}{\pi b} \arctan \left(\frac{\pi}{l}\right),
\end{eqnarray}
where we have denoted $t = \ln (m^2/\Lambda^2),~l_0 = \ln (\mu^2/\Lambda^2)$ and $l = \ln (k^2_{\perp}/(\beta\Lambda^2))$.
We have introduced the IR cut-off $\mu$ $(\mu \gg \Lambda)$
to guarantee the use of the perturbative QCD in Eq.~(\ref{aeff}).
The expression for $\alpha_s^{eff}$ in Eq.~(\ref{aeff}) was obtained in Ref.~\cite{etalpha} by applying the
Cauchy theorem to the evolution equation of the Bethe-Salpeter type for the Compton scattering amplitude.
Let us notice that such equations for the scattering amplitudes contain
$\alpha_s$ in the same way as in Eq.~(\ref{bsa})  (see Ref.~\cite{etalpha} for detail), so the
treatment of $\alpha_s$ in the equations for the structure functions (where the $s$-channel partons are on-shell) and
amplitudes (with off-shell partons in the $s$ -channel) is identical.
The term $\sim \delta (m^2 - \mu^2)$ in Eq.~(\ref{aeff}) corresponds to the cut of the bare gluon propagator $(m^2 + \imath \epsilon)^{-1}$ in Eq.~(\ref{bsa}) while the integral
corresponds to cutting the running coupling $\alpha_s (m^2)$.
The main contributions in the $\mu$-dependent terms in Eq.~(\ref{aeff}) cancel
each other at
\begin{equation}\label{mupi}
\mu^2 \gg \Lambda^2 e^{\pi} \approx 23 \Lambda^2
\end{equation}
 and after that the effective coupling is reduced to
 the standard expression:
\begin{equation}\label{aeffbigmu}
\alpha_{eff} = \alpha_s (k^2_{\perp}/\beta)[1 + O(\alpha_s)] \approx \alpha_s (k^2_{\perp}/\beta)
\end{equation}
which looks IR-stable, however as a matter of fact it depends on $\mu$ implicitly through Eq.~(\ref{mupi}).

\section{Theoretical grounds for using the fixed couplings}

Quite often the running QCD coupling is approximated by a fixed coupling. The motivation is obvious:  accounting for the
running coupling effects in the Feynman graphs or in evolution equations makes calculations much more involved. In order to
simplify them, $\alpha_s$ is kept fixed and the scale is specified a posteriori. In this Sect. we present a new approach to
introduce the fixed coupling and specify its scale. In order to make our forthcoming reasoning clearer let us first consider approximations of a simple mathematical expression

\begin{equation}\label{intfggen}
V = \int_a^b dx g(x) f (x)
\end{equation}
for different functions $f$ and $g$ regular in the integration region. In what follows we will associate
$g(x)$ with running $\alpha_s$ while $f(x)$ includes a generic notation either for DL or SL set of variables.
We assume for simplicity that $f(x), g(x) \geq 0$. Let us consider several ways to approximate $V$ by expressions
where $g(x)$ is not integrated together with $f(x)$. These approaches are known from the standard  mathematical
analysis.

\textbf{Approximation (A)} In the first place we consider the case when
function $f(x)$ in the interval $[a,b]$ varies much faster than $g(x)$. To some extent it corresponds
to calculations in DLA. In this case
$g(x)$ can be approximated by its fixed value and therefore

\begin{equation}\label{intfga}
V \approx  V_A = g(x_0)\int_a^b dx f (x),
\end{equation}
with $a \leq x_0 \leq b$.

\textbf{Approximation (B)} The functions $g(x)$ and $f(x)$ vary similarly in the interval $[a,b]$.
It corresponds to  calculations in SLA.
In this case the approximation of Eq.~(\ref{intfga}) becomes groundless and should not
be used. Moreover, in this case any approximation of $g(x)$ by any fixed value is totaly unreliable.
Nevertheless, if there is really need for it, one can use the
following approximation:

\begin{equation}\label{intfgb}
V \approx  V_B = \bar{g} \int_a^b dx f (x),
\end{equation}
where $\bar{g}$ is the averaged value of  $g(x)$:

\begin{equation}\label{gav}
\bar{g} = \frac{1}{b-a} \int_a^b dx g (x).
\end{equation}
In contrast to $g(x_0)$ of Eq.~(\ref{intfga}), the averaged value $\bar{g}$ in Eq.~(\ref{intfgb}) is an
integral characteristic of the behavior of $g(x)$ in the interval $[a,b]$.
%
%
%
%

\section{Frozen coupling for Double-Logarithmic Approximation}

In the Double-Logarithmic Approximation (DLA), each loop integration involves the running $\alpha_s$ accompanied by a set of
variables yielding  two logarithms (e.g. $(\sim \alpha_s \ln^2 s)^k$),
possibly with different arguments, when the coupling is fixed. DLA for Eq.~(\ref{bsaeff}) means that the blob $M$
 depends on $\beta$ and $k_{\perp}$ through logarithms only. As the running effective coupling
 $\alpha^{eff}_s$ in Eq.~(\ref{bsaeff}) involves single logarithmic
contributions, it
can  be regarded approximately as a slower varying term compared to the DL term. This makes possible to
use Approximation \textbf{(A)} and use in Eq.~(\ref{bsaeff}) the effective coupling

\begin{equation}\label{afixdl}
\alpha_s^{DL} (\mu) = \frac{1}{b} \frac{\ln (\mu^2/\Lambda^2)}{\left[\ln^2(\mu^2/\Lambda^2) + \pi^2\right]}
\end{equation}
given by the first term in Eq.~(\ref{aeff}), where $\mu$
 should be specified.
\subsection{Scale for the DL coupling}

In order to specify $\mu$ in Eq.~(\ref{afixdl}), we use the Principle of Minimal Sensitivity suggested in Ref.~\cite{pms}.
Since the early applications in Ref.~\cite{pms1}, this approach has been used to solve various QCD problems. For example,
in the context of the BFKL NLO Pomeron\cite{bfklnlo} it was used to set
the scale of the coupling in Ref.~\cite{kim} and applied to the analysis of electro-production of light vector mesons
in Ref.~\cite{iv}.
With regard of Eq.~(\ref{afixdl}), this approach
means that the optimal scale $\mu_0$ can be defined by the requirement

\begin{equation}\label{muopteq}
\frac{d \alpha_s^{DL} (\mu_0)}{d \mu} = 0.
\end{equation}
Obviously, the $\mu$-dependence of $\alpha_s^{DL} (\mu)$ is weaker at $\mu=\mu_0$ than at any other
values where the derivative is not equal to zero. Solving Eq.~(\ref{muopteq}) leads to the
DL optimal scale

\begin{equation}\label{muopt}
\mu_0 = \Lambda e^{\pi/2}
\end{equation}
and allows us to define the frozen coupling $\alpha_s^{DLT}$ at the time-like argument as

\begin{equation}\label{afrozss}
\alpha_s^{DLT} \equiv \alpha_s^{DL} (\mu_0) = \frac{6}{11 N - 2 n_f}.
\end{equation}
This coupling mimics the running coupling
in the evolution equations
of the Bethe-Salpeter type with the DL accuracy.
We have considered here the case when the factorized gluon propagates in the $s$ -channel.
In
other words, $\alpha_s^{DLT}$ is related to the ladder Feynman graphs.
However, there are alternative situations when the argument of $\alpha_s$ is space-like.
For example, in the first place, the ladder graphs do not yield DL contributions in the hard kinematics where
$s \sim -t \sim -u$. Besides, in some important processes (for example, in
the $e^+ e^-$ -annihilation and in the Drell-Yan process) DL contributions are of the
Sudakov type,  they come from non-ladder graphs where the gluons are
soft and the arguments of the
couplings $\alpha_s = \alpha_s (k^2_{\perp})$ are space-like, so the couplings do not have $\pi^2$ -terms.
In this case the expression for the fixed effective coupling is

\begin{equation}\label{afixu}
\widetilde{\alpha}_s^{DL} = \frac{1}{b \ln (\mu^2/\Lambda^2)}.
\end{equation}

Obviously, PMS cannot be applied straightforwardly to Eq.~(\ref{afixu}) to settle the scale. However,
the couplings in Eqs.~(\ref{afixdl})
and (\ref{afixu}) are related by the analyticity. So, by
substituting the scale $\mu_0$ of Eq.~(\ref{muopt}) in  Eq.~(\ref{afixu}) allows us to define the frozen
coupling $\alpha^{DLS}_s$ at the space-like argument:

\begin{equation}\label{afrozsu}
\alpha^{DLS}_s= \widetilde{\alpha}_s^{DL} (\mu_0) = 2 \alpha_s^{DLT} = \frac{12}{11 N - 2 n_f}.
\end{equation}

Now let us present the numerical values of the frozen couplings $\alpha_s^{DLT}$ and $\alpha_s^{DLS}$. It follows from
Eqs.~(\ref{afrozss},\ref{afrozsu})  that
at $n_f = 3$ $\alpha_s^{DLT} = 2/9 \approx 0.22$  and $\alpha_s^{DLS} = 4/9 \approx 0.44$.
When $n_f = 4$, the both couplings are slightly greater: $\alpha_s^{DLT}= 6/25 = 0.24$ and $\alpha_s^{DLS}= 12/25 = 0.48$.
Let us notice here that the estimates for $\alpha_s^{DLS}$ quite satisfactorily
agree with the early estimates in the $e^+ e^-$ -annihilation (see e.g. Ref.~\cite{greco}).

\section{Frozen coupling for Single-Logarithmic Approximation}

In contrast to the case of DLA considered in Sect.~IV,
there are situations where SL terms (e.g. the terms $\sim (\alpha_s \ln s)^k$) become the leading ones.
For instance, this takes place when DL contributions of different Feynman graphs cancel
each other.
The well-known example of such situation is the BFKL Pomeron
in the LO\cite{bfkl} and NLO\cite{bfklnlo}
where logarithms of $k_{\perp}$ coming from different
graphs cancel each other, so that the remaining single logarithms are logarithms of $\beta$.
In this case keeping the QCD coupling fixed and accounting for other
single-logarithms has no real theoretical grounds.
Nevertheless, it was done in Refs.~\cite{bfkl,bfklnlo} while the scale of the coupling was
fixed a posteriori in Ref.~\cite{kim}. Below we suggest alternative ways to fix the coupling
in SLA,  according to
Eq.~(\ref{intfgb}).   Similarly to Sec.~IV, we consider separately the cases of the couplings at time-like and
space-like arguments. In the first place, we focus our consideration on the coupling used in
BFKL,
i.e. the couplings in the gluon ladders where their arguments are time-like.

\subsection{Scale for the SL coupling for  time-like arguments}

We start by considering  Eq.~(\ref{aeff}) where $l_0$ satisfies Eq.~(\ref{mupi}). So, we assume that

\begin{equation}\label{lz}
l_0 = z \pi,
\end{equation}
with the minimal value $z^2_{min} \sim 10$. In this case the $l_0$ -depending terms in Eq.~(\ref{aeff})
can be dropped, so

\begin{equation}\label{aeffbigl}
\alpha_{eff}  \approx
\frac{1}{\pi b} \arctan \left(\frac{\pi}{l}\right) =
\frac{1}{\pi b} \arctan \left(\frac{1}{y}\right),
\end{equation}
with $y = l/\pi$.

Let us define the averaged integral coupling $<\alpha_{eff}>$  as

\begin{equation}\label{aavtent}
<\alpha_{eff}> = \frac{1}{l} \int dl \alpha_{eff} (l) \approx
\frac{1}{\pi^2 b y} \left[1 + \ln y\right] = \frac{0.16}{y} \left[1 + \ln y\right] .
\end{equation}
The substitution of the minimal possible value for $y$, $y \sim z_{min} = \sqrt{10}$ into (\ref{aavtent})
yields that the maximal value of $<\widetilde{\alpha}_{eff}>$:

\begin{equation}\label{asltent}
\max[<\alpha_{eff}>]  \equiv \alpha^{SLT}_s \approx 0.11.
\end{equation}


%
%
%
%
%
Therefore, the direct analysis of the effective QCD coupling has not made possible
to settle the scale of the single-logarithmic frozen coupling for  time-like arguments. However, it allows us to fix  in Eq.~(\ref{asltent}) the upper value
$\alpha_s^{SLT} \approx 0.1$ of
the effective SL coupling.

\subsection{Scale for the SL coupling for space-like arguments}

It is also important to estimate the single-logarithmic coupling for space-like arguments. For instance, this
happens in the graphs where the virtual quark blobs are attached to the external quarks by $t$-channel gluons.
An example of such  graph contributing to the photon-quark forward scattering is depicted in Fig.~3 where $\alpha_s = \alpha_s (k^2)$
and $k^2 < 0$:

\begin{figure}[h]
\begin{center}
\begin{picture}(125,160)
\put(0,0){ \epsfbox{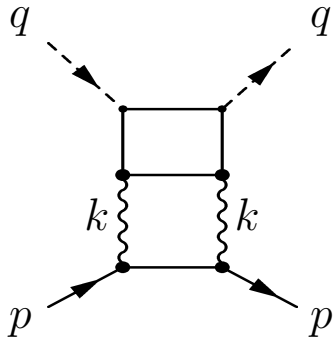} }
\end{picture}
\end{center}
\caption{\label{frozenfig3} An example of the graphs with the single-logarithmic
coupling at the space-like arguments.}
\end{figure}

 Using the perturbative expressions for $\alpha_s$ for this graph and integrating over $k^2$
 (at high energies $k^2 \approx - k^2_{\perp}$), with
the integration running from $k^2_{\perp} = 0$  to $k^2_{\perp} = \infty$, inevitably leads to
infrared (IR) divergence at $k^2_{\perp} = 0$. In order to regulate it, one has to introduce an IR cut-off.
For example, it can be done either with the restriction of the transverse phase space like $k^2_{\perp} > \mu^2$ or with the shift
$k^2_{\perp} \to k^2_{\perp} + \mu^2$, which leads to the shift

\begin{equation}\label{shifta}
\alpha_s (k^2_{\perp}) \to \alpha_s (k^2_{\perp} + \mu^2)
\end{equation}

and therefore,  $\alpha_s$ tends to some frozen value $ \equiv \alpha_s (\mu^2_{SLS})$ when $k^2_{\perp} \to 0$. Obviously, the leading
contribution of the integration
over $k$ in Fig.~3 is single-logarithmic whereas the DL contributions are suppressed by a power of the
invariant energy $2pq$. This feature keeps unchanged when more the $t$-channel (vertical) gluon propagators
are inserted into Fig.~3 to connect the quark blob to the external quarks.
So, similarly to the reasoning of above, we will use again
Eq.~(\ref{gav}) in order to fix the scale of $\mu$ in Eq.~(\ref{shifta}). This time the
averaged coupling is

\begin{equation}\label{aavs}
<\alpha_s> = \frac{1}{l} \int dl' \alpha_s (l') = \frac{1}{b} \frac{\ln l}{l} =
\frac{1}{b \pi} \frac{\ln\pi y}{ y},
\end{equation}
where we have used the notations of Eqs.~(\ref{lz},\ref{aeffbigl}). Choosing
$y \approx \sqrt{10}$, as we did it in  Eq.~(\ref{aeffbigl}),we arrive at the estimate:

\begin{equation}\label{asls}
\alpha_s^{SLS} \equiv  \frac{1}{b \ln (\mu^2_{SLS}/\Lambda^2)}= \max [<\alpha_s>] \approx \frac{1}{4.33 b}
\end{equation}

 which is about a factor of 3 larger than the time-like case. Therefore our estimate for the scale $\mu_{SLS}$ corresponding to $\alpha_{SLS}$ at $n_f =3$ is
$\mu_{SLS} = 8.7 \Lambda$. To conclude this Sect., let us stress once more that treating the
QCD coupling as a fixed coupling in the context of the SLA is dubious and, strictly speaking,
not supported by any theoretical ground. Nonetheless, although
we were not able to evaluate the values of the fixed couplings in the SLA, we have performed  reasonable
estimates for their maximal values, $\alpha_s^{SLT}$ and $\alpha_s^{SLS}$, for the time-like and space-like
arguments respectively.


\section{Check on the PMS predictions for the frozen couplings $\alpha_s^{DLT}$ and $\alpha_s^{DLS}$}

The values of the frozen double-logarithmic couplings $\alpha_s^{DLT}$ and $\alpha_s^{DLS}$
were obtained in Sect.~V by exploiting PMS.  However,
the results obtained with PMS can be regarded rather as suggestions than strict predictions. So, before
using our results for the values of $\alpha_s^{DLT}$ and $\alpha_s^{DLS}$, we prefer to perform some checks on them.
By doing so, we would like to avoid any direct comparison to
experiments because the experimental data usually involve many phenomenological
parameters in addition to the couplings. Fortunately, in the available literature there are results free of phenomenological parameters
but involving these couplings: the intercepts of the DIS structure functions. The point is that
the total resummation of the leading logarithmic ( either Dl or SL) contributions to the DIS structure functions leads to their Regge
asymptotics $ x^{- \Delta}$, with $\Delta$ being defined as the intercepts.  The Regge asymptotics using the
approximation of the fixed coupling
was obtained in Refs.~\cite{bfkl, bfklnlo} for the singlet
$F_1$ (resummation of SL contributions depending on $x$) and in Ref.~\cite{emr,bersns} for $F_1^{NS}$
and $g_1^{NS},~g_1^{S}$ (total resummation of DL contributions regardless of their arguments). The QCD coupling in
those papers was kept fixed and its scale was specified a posteriori. The scale for
$\alpha_s$ in the expressions for the BFKL Pomeron intercepts was settled in Ref.~\cite{kim}.
In the present Sect. we estimate those intercepts, using the numerical values of the fixed couplings obtained in the
previous Sects., and then compare them to alternative calculations performed without exploiting the approximation of frozen coupling.

\subsection{Non-singlet intercepts at the frozen couplings}

We consider here the intercepts of the structure functions $F_1^{NS}$ and $g_1^{NS}$
calculated in Refs.~\cite{emr, bersns} in DLA.
The intercept of $F_1^{NS}$
in DLA involves the frozen coupling $\alpha_s^{DLT}$ only:

\begin{equation}\label{intfns}
\Delta_{FNS}^{DL} = \sqrt{2 \alpha_s^{DLT} C_F/\pi}.
\end{equation}
In contrast,  the intercept of $g_1^{NS}$ involves both $\alpha_s^{DLT}$ and
$\widetilde{\alpha}_s^{~sud}$:

\begin{equation}\label{intgns}
\Delta_{gNS}^{DL} \approx \Delta_{FNS}^{DL}\sqrt{\frac{1}{2}\left[1 + \sqrt{1 +
\frac{\alpha_s^{DLS}}{\alpha_s^{DLT}} \frac{4}{(N^2 - 1)}}\right]}.
\end{equation}

Substituting the numerical values of the frozen couplings\footnote{The calculations actually made in Ref.~\cite{bersns}
involved only one coupling treated as a fixed parameter.} into Eqs.~(\ref{intfns},\ref{intgns}),
leads to the estimates  $\Delta_{FNS}^{DL} = 0.43$ and $\Delta_{gNS}^{DL} \approx 0.47$.

\subsection{Accuracy of the use of PMS in Double-Logarithmic Approximation}

The intercepts of $F_1^{NS} (x,Q^2)$
and $g_1^{NS} (x,Q^2)$ were calculated in Ref.~\cite{egtint} with the
running $\alpha_s$ effects accounted for. The numerical values of the intercepts were:
$\Delta_{FNS} = 0.38$ and $\Delta_{gNS} = 0.42$, so the discrepancy are:
$(|\Delta_{FNS}^{DL} - \Delta_{FNS}|)/\Delta_{FNS} \approx 0.13$ and
$(|\Delta_{gNS}^{DL} - \Delta_{gNS}|)/\Delta_{gNS} \approx 0.12$.
This comparison  allow us to conclude that \\
\textbf{(i)} The use of the frozen couplings leads to a greater values of the intercepts than the use of the running
coupling \\
\textbf{(ii)} The accuracy can be estimated  to be   within $15 \%$ approximately.

\section{Check of our estimates for the SL couplings}

Unfortunately,  no calculations of the SL intercepts with running coupling effect taken into account,
 are present  in the literature, contrary to the DL case. This makes impossible a straightforward check on our estimates for the
single-logarithmic frozen couplings $\alpha_s^{SLT}$ and $\alpha_s^{SLS}$. The only intercept
we can use for the check is the NLO BFKL intercept. This
is approximately given
by the following expression\cite{bfklnlo}:

\begin{equation}\label{intnlo}
\Delta^P_{NLO} \approx \Delta^P_{LO} \left(1 - B \alpha_s\right) \equiv A \alpha_s^{BFKL} \left(1 - B \alpha_s^{BFKL}\right),
\end{equation}
with $A = 4N \ln 2/\pi\approx 2.65$ and $B \approx 6.36$. The scale of the coupling $\alpha_s^{BFKL} \approx 0.1$ was found in
Ref.~\cite{kim}. Using it in Eq.~(\ref{intnlo}) led to to the value of Pomeron intercept
$\Delta^P_{NLO} = 0.08$. Using the estimate $\alpha_s^{SLT}$ of Eq.~(\ref{asltent}) yields a quite similar result:

\begin{equation}\label{inttilde}
\widetilde{\Delta}^P_{NLO} \approx 0.087 \approx 0.09.
\end{equation}
Comparison of our estimate in Eq.~(\ref{inttilde}) to the estimate 
obtained in Ref.~\cite{kim} leads to $[\widetilde{\Delta}^P_{NLO} - \Delta^P_{NLO}]/\Delta^P_{NLO} = 0.12$.

The value of $\alpha_s^{BFKL}$ in Ref.~\cite{kim} was obtained with several technical means, including the use of PMS. It is interesting
to notice that the straightforward application of PMS to Eq.~(\ref{intnlo}) yields the results quite compatible
with both the result of  Ref.~\cite{kim} and our estimate in Eq.~(\ref{inttilde}). Indeed,
application of PMS to Eq.~(\ref{intnlo}) means
solving the equation
\begin{equation}\label{pmsdeltap}
\frac{d \Delta^P_{NLO}}{d \alpha_s} = A(1 -2B \alpha_s) =0.
\end{equation}
It yields the following value for the frozen SL coupling:
\begin{equation}\label{aslpms}
\alpha_{PMS}^{SL} = 1/2B \approx 0.08 .
\end{equation}

Substituting
$\alpha_{PMS}^{SL}$ into Eq.~(\ref{intnlo}) leads to

\begin{equation}\label{intpms}
\Delta^P_{PMS} = 2.65 \alpha^{SL}_{PMS} \left(1 - 6.36 \alpha_{PMS}^{SL}\right) \approx 0.1.
\end{equation}

The substitution of the couplings $\widetilde{\alpha}^{SL} = 0.1$  into Eq.~(\ref{intnlo})
leads to the same values of the intercepts: $\Delta^P_{LO} \approx 0.21$ and $\Delta^P_{NLO} \approx 0.1$ respectively.

\section{Comment on the $Q^2$ -dependence of the QCD coupling}

The problem of the frozen coupling is often related to the replacement\footnote{the shift in Eq.~(\ref{shiftq})
should not be confused with the shift in Eq.~(\ref{shifta}).}
\begin{equation}\label{shiftq}
\alpha_s (Q^2) \to \alpha_s (Q^2 + \mu^2)
\end{equation}
in order to prevent the infrared divergence of
the coupling the region of small $Q^2$. In Eq.~(\ref{shiftq}) the
notation $Q^2$ stands for an external virtuality (for instance, it denotes the virtuality
of the external photon in the context of DIS structure functions).
To this regard we remind that the parametrization
$\alpha_s = \alpha_s (Q^2)$
is an approximation originally
introduced in the framework of DGLAP at large $x$ and large $Q^2$. This parametrization
follows from Eq.~(\ref{aeff}) (see Ref.~\cite{etalpha} for detail) when $\beta$
is $\sim 1$, the cut-off $\mu$ satisfies Eq.~(\ref{mupi}) and the upper limit of the
$k_{\perp}$ -integration in Eq.~(\ref{bsa}) is $Q^2$ (i.e. when $x$ is not far from $1$
and therefore the invariant energy can be approximated by $Q^2$).
However, this
parametrization has often been used ad hoc
in the region of small $x$ and small $Q^2$, where the theoretical basis of
DGLAP fails.
Moreover, sometimes the parametrization $\alpha_s (Q^2)$ is used  in the expressions for
the small-$x$ asymptotics (though this kinematic region is far beyond the applicability region of DGLAP),
leading to intercepts depending on $Q^2$.
We stress that this parametrization should never be regarded as a rigorous expression
where the value of $Q^2$ could be arbitrary.
In Ref.~\cite{egtint} (see also overviews \cite{acta}) we have demonstrated that
using the parametrization $\alpha_s (Q^2)$ in the small-$x$ region leads to a
wrong small-$x$ behavior even for the simplest, non-singlet structure functions.
To conclude this remark, we would like to remind that the $Q^2$ -dependence of the small-$x$
asymptotics of the structure functions is not related to $\alpha_s (Q^2)$ at all.
Indeed, the small-$x$ asymptotics of the DIS structure functions is universal and given by
the following expression:

\begin{equation}\label{as}
f \sim x^{-\Delta} \left(Q^2/Q^2_0\right)^{\Delta/2},
\end{equation}
with the intercept $\Delta$ being a number (of course, different structure functions have different intercepts). Eq.~(\ref{as}) means that
the asymptotic scaling takes place at small $x$ and large $Q^2$:
the DIS structure functions depend only on one argument $\left(Q^2/\sqrt{x}\right)$.
 This scaling is not sensitive on whether the coupling is running or fixed.

\section{Summary}

In the present paper we have suggested a novel way to define the frozen QCD couplings and
fix their value. The definition of a  frozen couplings
comes from the analysis of the effective coupling in the evolution
equations. It depends on whether the leading logarithms are double- or single-
logarithmic. The value of the DL couplings has been fixed by applying the Principle of Minimal Sensitivity. We
have  found a value $0.24$ for the DL coupling for a time-like argument (non-singlet structure functions and
scattering amplitudes in the forward kinematics with a non-vacuum quantum numbers in the $t$ -channel) and $0.48$
 for a space-like argument (e.g. for the scattering amplitudes in the hard or backward kinematics, $e^+ e^-$ -annihilation
 and DY processes). In contrast, our result for the values of the SL couplings
contributing to the BFKL Pomeron intercept and other important reactions is less rigorous. Our estimate for the value
of SL coupling at time-like argument is $\approx 0.1$ and when the argument is space-like, the coupling is $0.3$.
In order to check our estimates, we have compared them to the results available in the
literature and at the same time independent of other phenomenological parameters.
By doing so, we have considered the intercepts of DIS
structure functions. The comparison shows that the use of
our frozen DL couplings for  the intercepts of the non-singlet structure functions
differs from the more accurate calculations\cite{egtint} (which  account for the running coupling
effects) by  $13\%$. On the other hand the discrepancy between our estimate for
the intercept of BFKL Pomeron and the results of Ref.~\cite{kim}  (where an alternative setting of the scale
is used) is also within $15\%$.

\section{Acknowledgements}

We are grateful to S.I.~Alekhin, D.~Colferai, V.T.~Kim, A.V.~Kotikov, G.I.~Lykasov, N.N.~Nikolaev,
V.A.~Petrov, B.R.~Webber, B.G.~Zakharov and N.P.~Zotov for useful correspondence.
The work is partly supported by Grant RAS 9C237,
Russian State Grant for Scientific School RSGSS-65751.2010.2 and
EU Marie-Curie Research Training Network under contract
MRTN-CT-2006-035505 (HEPTOOLS).

\end{document}